# Automated Mechanism to Support Trade Transactions in Smart Contracts with Upgrade and Repair

Christian Liu
Faculty of Computer Science
Computer Science
Halifax, Canada
Chris.Liu@dal.ca[1]

Peter Bodorik
Faculty of Computer Science
Computer Science
Halifax, Canada
Peter.Bodorik@dal.ca

Dawn Jutla
Sobey School of Business
Saint Mary's University
Halifax, Canada
Dawn.Jutla@gmail.com

**Abstract:** In our previous research, we addressed the problem of automated transformation of models, represented using the business process model and notation (BPMN) standard, into the methods of a smart contract. The transformation supports BPMN models that contain complex multi-step activities that are supported using our concept of multi-step nested trade transactions, wherein the transactional properties are enforced by a mechanism generated automatically by the transformation process from a BPMN model to a smart contract. In this paper, we present a methodology for repairing a smart contract that cannot be completed due to events that were not anticipated by the developer and thus prevent the completion of the smart contract. The repair process starts with the original BPMN model fragment causing the issue, providing the modeler with the innermost transaction fragment containing the failed activity. The modeler amends the BPMN pattern on the basis of successful completion of previous activities. If repairs exceed the inner transaction's scope, they are addressed using the parent transaction's BPMN model. The amended BPMN model is then transformed into a new smart contract, ensuring consistent data and logic transitions. We previously developed a tool, called TABS+, as a proof of concept (PoC) to transform BPMN models into smart contracts for nested transactions. This paper describes the tool TABS+R, developed by extending the TABS+ tool, to allow the repair of smart contracts.

Keywords: Blockchain, Smart contracts, BPMN, Transaction mechanism, Automated generation, Smart contract upgrade or repair, TABS+R

## 1. Introduction

The publication of the Bitcoin white paper in 2008 and the launch of the Bitcoin blockchain in 2009 have sparked significant interest and research into blockchain technology. This technology has garnered attention from businesses, researchers, and the software industry due to its appealing characteristics, such as trust, immutability, availability, and transparency. However, like any emerging technology, blockchains and their smart contracts introduce new challenges, particularly concerning blockchain infrastructure and smart contract development.

Researchers are actively addressing several key issues, such as blockchain scalability, transaction throughput, and high costs. For instance, the high cost associated with consensus algorithms has been thoroughly studied, leading to the development and implementation of new consensus mechanisms. Additionally, challenges specific to smart contract development, such as limited stack space, the oracle problem (the blockchain's inability to interact with external data), data privacy, and compatibility across different blockchains, have also been explored in depth. Comprehensive literature reviews on these topics are available from various sources [1-10].

The constraints imposed by blockchains increase the complexity of smart contract development, especially for distributed collaborative applications. This complexity is highlighted by numerous literature surveys on the topic, such as those by Taylor (2019) [2], Khan (2021) [3], Vacca (2021) [4], Belchior (2021) [5], Saito (2016) [6], Garcia-Garcia (2020) [7], Lauster (2020) [8], and Levasseur (2021) [9]. To simplify smart contract development, studies by López-Pintado (2019) [9, 11], Tran (2018) [12], Mendling (2018) [13], and Loukil (2021) [14] propose to express the requirements of a blockchain application using a model expressed in the business process model and notation (BPMN), which is then transformed into a smart contract.

Our research also starts with a BPMN model that is automatically transformed into smart contract methods, but our approach differs significantly as we use multi-modal discrete event hierarchical state machine (DE-HSM) modeling to transform the BPMN model into a DE-HSM model that allows for graph-based representations of distributed blockchain applications, facilitating analysis and identifying patterns that remain isolated from other concurrent activities. We describe our approach in Refs. [15-22], together with a TABS tool (Tool to Automatically Transform a BPMN Model to Smart Contract Methods) that we developed as a proof of concept (PoC) to demonstrate the feasibility of our approach. In Ref. [22], we expand our approach presented in Ref. [15] to address collaborative activities in trade and distributed finance, to which we refer simply as trade activities. These activities are often performed by several participants executing various multi-step activities, such as price negotiations, letters of credit, transportation, and exchanges of various documentation. A smart contract naturally represents such collaborative activities employing several methods, and synchronization of such activities is thus required.

---

[1] Corresponding author

However, a native blockchain transaction often falls short of representing these complex trade activities due to its focus on state changes rather than the collaborative nature of trade transactions. We use the term native blockchain transaction to refer to the general concept of a blockchain transaction. If a blockchain supports native cryptocurrency, we consider any transfer of a native cryptocurrency as a part of native blockchain transaction. The problem is that a trade transaction is naturally expressed as a collaboration of several methods that are invoked independently by the participants of trade activities, wherein a native blockchain transaction supports only the concept of a transaction made by an execution of any of the methods of a smart contract (which of course may invoke other methods). The native blockchain transactions thus cannot include updates to the ledger made by two smart contract methods that were independently invoked by the distributed application. This mismatch is similar to the object-relational impedance mismatch [23]. We addressed this issue in our previous research [20] by proposing a methodology that allows developers to define a multi-step trade transaction, simply referred to as a trade transaction, as a collection of smart contract methods that can be invoked by different trade participants. We adapted database transactional properties (Atomicity, Consistency, Isolation, and Durability, or ACID) for trade transactions and incorporated features to provide access control and privacy. Our approach uses pattern augmentation techniques to automate the creation of mechanisms that enforce these properties.

To develop a smart contract involving trade transactions, the developer writes the methods as usual and identifies the methods forming the trade transaction, and the transformation process from BPMN to smart contracts uses our methodology to support the multi-step trade transaction properties. We also support nested trade transactions, while imposing some restrictions to ensure that trade transaction methods refer only to objects and methods within their defined scope. Since our initial proposal in Ref. [17], we have integrated nested trade transactions into our automated BPMN-to-smart-contract transformation project [22], exploring mechanisms to support transactional properties and their impact on access control, privacy, and recovery. However, recognizing the importance of handling exceptions, we shifted our focus to automating recovery procedures for trade transactions, as smart contracts often encounter failure scenarios.

A trade transaction, which is multi-step and may be nested, involves the execution of multiple methods of a smart contract, and thus recovery from a failure is more complex than recovery from a failure of a native blockchain transaction. A recovery procedure needs to ensure that (i) the ledger is not affected by a failed transaction, and that (ii) different actors that participate in the trade transaction execution are informed of the failures in the correct sequence so that they can recover their resources dedicated to the execution of the failed trade transaction on their local systems.

In addition to invoking recovery procedures for the application, we also address the issue of the failure of smart contracts when real-life situations prevent their completion. In real life, if trade activity arrangements cannot be completed due to some events or conditions, alternative arrangements are made. However, if such trade activities are modeled by a smart contract, the question is how to update or repair the smart contract to represent the trade activity with new alternative arrangements.

Software application life cycle includes upgrades to fix bugs and introduce new features to respond to new requirements caused by ever-changing environment. Smart contracts are not any different, but owing to the blockchain immutability, upgrading smart contracts causes difficulties, with active research addressing the problem, as judged by surveys on the topic [50-52]. However, real life complicates issues even further. A situation may arise in which a trade activity cannot be completed using the original arrangements due to some unanticipated events or conditions, and new arrangements need to be made. Thus, a smart contract developed to represent the original activities needs to be upgraded to represent the newly arranged activities to facilitate their successful completion. Thus, not only does the smart contract need to be upgraded, but, if possible, the upgrade should avoid redoing completed activities, and we refer to such an upgrade as a smart contract repair. The question arises as to how to repair the activities of a smart contract while retaining the partially completed activities and ensuring consistency, which we also address in this paper.

### 1.1. Objectives and contributions

In our previous research [15, 22], we addressed the issue of generating smart contracts from BPMN models with the support of nested transactions, defined over a subset of methods of a smart contract, to support multi-step trade transactions performed by several transaction participants. Developers declare a trade (sub)transaction as a collection of methods, while the automated transformation from a BPMN model to the methods of smart contracts also provides an automated transaction mechanism to support the multimethod and possibly nested transactions.

In this paper we describe recovery procedures for the multi-step trade transactions and our approach to the smart contract repair. Our approach not only supports recovery from failure but also facilitates repair of a smart contract: if the execution of a trade activity, as represented by the smart contract, fails due to an unanticipated situation, alternative arrangements are made in order to complete the trade. Such alternative arrangements strive to reuse already completed trade activities in order to reduce the overall cost. However, the smart contract also needs to be upgraded to represent the new alternative arrangements that also avoid recovering and redoing activities that have been already completed successfully by the original smart contract before its failure exception occurred.

The specific objectives, which also constitute the contributions of this paper, are as follows:
- **Objective 1**: Describe the process for recovering a failed trade (sub)transaction to the state just before the transaction begins. This recovery includes not only the restoration of the transaction on the blockchain, but also the invocation of recovery procedures for the distributed application, enabling it to recover local resources dedicated to the processing of the failed (sub)transaction.
- **Objective 2**: Investigate a methodology for trade (sub)transaction repair with the following considerations:
  - If possible, ensure that trade (sub)transactions representing successfully completed trade activities remain unaffected.
  - If possible, ensure that unexecuted trade (sub)transactions representing trade activities that follow the repaired/amended trade activity remain unaffected.
- **Objective 3**: Develop a PoC that demonstrates the feasibility of the proposed methodology for transaction repair.

To briefly summarize our contributions, we present our initial approach to repairing trade (sub)transactions in smart contracts to reflect alternative arrangements. We utilize nested trade transactions to facilitate repairs, focusing on amending only the failed (sub)transaction rather than the entire trade activity. We automate the generation and deployment of smart contracts and describe how to create and update versions of failed (sub)transactions, ensuring continued execution of the smart contract post-repair. We describe a tool, called TABS+R, which we developed as a PoC.

## 1.2. Outline

Section 2 provides the necessary background. Section 3 details the recovery process for failed trade (sub)transactions, which restores the system to the state prior to the transaction's invocation. This process must ensure that:
1. The ledger state remains unaffected by the failed (sub)transaction.
2. Recovery procedures for transaction participants are triggered to release local resources allocated for the failed (sub)transaction.

To address real-life scenarios where a trade activity failure, represented by a trade transaction, which requires amendments, Section 4 describes our approach to repairing trade transactions on the basis of the structure of nested trade transactions within a smart contract generated from a BPMN model. We explain how trade transactions are encapsulated in separate smart contracts and describe the process of repair by replacing the smart contract for the failed (sub)transaction with a revised version. The section also discusses the constraints and implications of such repairs on any preceding or subsequent activities related to the failed (sub)transaction.

Section 5 describes modifications made to our tool, TABS+, to create a tool TABS+R that supports transaction repair as a PoC. It discusses the potential benefits of supporting trade transaction repairs and identifies obstacles that need to be overcome for the broad adoption of our approach to the automated generation of smart contracts with repair capabilities.

Section 6 provides an overview of related work in the field and discusses limitations. Finally, Section 7 presents the conclusions of the study and describes future research directions.

## 2. Background

This section first provides an overview of BPMN modeling and then discusses modeling with finite state machines (FSMs), hierarchical state machines (HSMs), and multimodal modeling. We also review transactions in database and blockchain systems, comparing their properties with our concept of trade transactions. These concepts are foundational for our approach to repairing trade transactions generated from BPMN models. As this research extends our previous work on the automated generation of smart contracts from BPMN models, the section overviews our approach to automatically generating smart contracts from BPMN models.

## 2.1. Business process model and notation (BPMN)

BPMN, developed by the object management group (OMG) [24-27], is designed to be comprehensible to a wide range of business users, from analysts to technical developers and managers. Its practical adoption is evident from various software platforms that enable modeling business applications with the goal of automatically generating executable applications from BPMN models. For example, the Camunda platform transforms BPMN models into Java applications [28], while Oracle Corporation converts BPMN models into executable process blueprints via the business process execution language (BPEL) [29].

Key features of BPMN models include flow elements that represent computation flows between BPMN elements. A task represents a computation executed when the flow reaches it. Other elements manage conditional forking and joining of computation flows, using Boolean expressions (guards) to control the flow or represent event handling. Additionally, data elements describe the data or objects flowing with the computations, serving as inputs for decision-making in guards or computation tasks.

2.2. FSMs, HSMs, and multimodal modeling

FSM modeling is widely used in software design and implementation and is often extended with features such as guards in FSM transitions. In the late 1980s, FSMs evolved into HSMs, incorporating hierarchical structures to facilitate pattern reuse, allowing states to contain other FSMs [30].

Girault et al. (1999) [31] described combining HSM modeling with concurrency semantics from models like communicating sequential processes [32] and discrete events [33]. They describe how a system state can be represented by an HSM, with a specific concurrency model, which is applicable only to that state. This supports multimodal modeling, where different hierarchical states can employ the most suitable concurrency models for concurrent activities in that state.

2.3. BPMN Model Transformation to Smart Contract Methods

In Refs. [15, 22], we presented a methodology for transforming BPMN models into smart contracts. The transformation process involves several key steps:

1. **Transformation of the BPMN model to directed acyclic graph (DAG) representation:** The BPMN is pre-processed and is converted into a DAG representation. The mapping ensures that for any DAG vertex or edge, the corresponding BPMN element can be identified, and vice versa.

2. **Identification of single-entry single-exit (SESE) subgraphs:** The DAG is analyzed to identify SESE subgraphs. A SESE subgraph is such that it has a single-entry vertex, i.e., the only vertex in the subgraph that has an input edge from a vertex outside the subgraph, and a single-exit vertex, i.e., the only vertex in the subgraph that has an output edge leading to a vertex outside the subgraph. All other subgraph vertices have only edges connected to the internal nodes of that subgraph. SESE subgraphs are significant because they represent the flow of computation, and once the computation flow, represented by the graph edges, enters a SESE subgraph, it remains confined within that subgraph until it exits via the subgraph's exit node. This ensures a contained and manageable flow of computation. The identified SESE subgraphs are shown to the developer who decides which of those subgraphs should be implemented by the transformation process as transactions. Any chosen SESE subgraph that contains other proper SESE subgraphs will be implemented as a parent transaction containing nested subtransactions, one for each of its SESE subgraphs, which is applied recursively.

3. **Transformation to discrete event-finite state machine (DE-FSM) model:** The DAG, with its identified SESE subgraphs, is transformed into a discrete event hierarchical state machine (DE-HSM) model. Each node in the DE-HSM model represents either a DE-HSM sub-model or a computation expressed using concurrent FSMs, with some FSM states indicating execution of BPMN tasks. The DE-HSM model is further detailed by elaborating each of the HSMs until the whole BPMN model is flattened into a network of DE-FSM sub-models.

4. **Transformation into the smart contract methods:** The interconnected DE-FSM models are then transformed into a smart contract code. Each BPMN task element is represented as a separate method within the smart contract. Task-method executions are triggered by specific state transitions in the FSMs, making the system's collaborative activities, i.e., the business logic, independent of the underlying blockchain infrastructure. Thus, the deployment of independently executed tasks can be managed separately from the blockchain layer.

A smart contract is essentially an execution engine for concurrent FSMs. That is, each BPMN element representing a BPMN task computation is transformed into a separate method of a smart contract. A task-method execution is triggered when an FSM state is reached, which indicates that on a transition to that state, a particular task should be executed. The collaborative activities are thus represented by state changes in concurrent FSMs and are hence independent of the blockchain infrastructure. Thus, only the execution of the independently executed BPMN tasks is blockchain dependent.

We exploited the concept of multimodal modeling and independent subgraphs in Ref. [5], and then in our subsequent work of the project, we support sidechain processing by enabling the developer to choose and deploy a SESE subgraph as a separate transaction that is deployed on a sidechain. Thus, if a SESE subgraph has much computation to perform, such computation can be performed on a sidechain, albeit at the cost of overhead for communication between the mainchain and a sidechain. If computation performed on a sidechain is much cheaper than on the mainchain, then sidechain processing may be beneficial.

In Ref. [22], we use the nested structure of the SESE subgraphs to define nested trade transactions, which were initially introduced in Ref. [19], in the context of automated transformation of BPMN models into the methods of a smart contract. The BPMN model is transformed into a DAG and then into the DE-HSM model, and the developer is provided with information to decide which SESE subgraphs will be deployed as trade (sub)transactions, wherein the system automatically generates the transaction mechanism for each trade (sub)transaction. We facilitate options for a developer to select how the trade (sub)transactions should be packaged and deployed, wherein one option packages and deploys each trade (sub)transaction as a separate smart contract. It is this option that is used to support the repair of smart contracts.

The transaction mechanism, to support the nested multistep trade transactions, is generated by the transformation process from a BPMN model to a smart contract using a pattern augmentation scheme [22]. Ledger writes are not applied to the ledger directly, but instead, are cached and then applied to the ledger only during the trade transaction commit phase after all ledger updates have been cached. Therefore, if a blockchain transaction fails, ledger recovery is unnecessary. However, participants must be informed of the failure so that they can release local resources allocated for the failed transaction's processing.

## 3. Recovery procedures for nested trade transactions

In this section, we discuss automated recovery procedures for nested trade transactions, focusing on restoring the system to the state just before the transaction failure. We will first outline the recovery procedures specific to blockchain transactions, followed by a discussion of how these procedures are facilitated within the context of the trade transaction framework.

When an exception occurs during the execution of a smart contract method, the system checks for an associated exception handler. If the developer has provided an exception handler in the smart contract, it is invoked. This handler may resolve the issue, allowing the transaction to proceed without requiring recovery. However, if the exception cannot be resolved by the handler, the trade transaction fails, necessitating the execution of recovery procedures. Blockchain transactions, including trade transactions, can fail due to exceptions during the execution of a smart contract or during the consensus phase, where the blockchain ensures consistency and serializability of transactions.

### 3.1. Recovery for trade transactions

The recovery procedure for native blockchain transactions is straightforward, as the blockchain infrastructure inherently ensures ACID properties. However, trade transactions involve multiple actors, each committing resources on their systems, thus complicating the recovery process.

When a trade transaction fails, the recovery procedure must ensure that:
1. The ledger state remains unaffected by the failed transaction's execution.
2. All participants are notified of the failure so that they can release locally committed resources.

Since the trade transaction mechanism commits ledger updates only during the successful commit phase, there is no need for ledger recovery. However, recovery procedures for participant applications must follow the reverse order of the invocation of trade transaction methods. For nested trade transactions, this means that the recovery procedures of subtransactions must be executed before those of the parent transaction.

### 3.2. Trade activities, nested transactions, and recovery procedures

Consider a simple example of a smart contract that supports a trade transaction for the sale of a large product, such as a combine harvester. It may include price negotiation with payment via an escrow account, which is then followed by arranging transport. Transport arrangements include finding the requirements for the transport of the product, such as wide-load requirements or safety requirements in the case of dangerous products in transport. Once the transport requirements are determined, the insurance and transport are arranged, and the product is shipped/transported. Following the transport, the product is received, and payments are completed. Fig. 1 shows the trade activities as a BPMN model created using the Camunda platform invoked from our TABS+R tool, which we describe in a later section. However, the model can also be viewed as a block diagram of the trade activities we use for exposition purposes. Fig. 1 represents the following trade activities:

- *PriceAndEscrow* includes price negotiation and escrow payment.
- *GetTrRequirements* includes determining the transport safety requirements.

- GetRailInsurance includes obtaining the rail insurance to cover the product transport while satisfying the safety requirements.
- GetRailTrnasporter includes hiring the company to transport the product while satisfying the safety requirements.
- DoTransport includes the actual transport of the product
- ReceiveAndFinalize includes customer's acceptance of the delivered product and completion of the payment.

Recall that the transformation process, from a BPMN model into the methods of a smart contract, uses the concept of SESE subgraphs to find BPMN patterns that are suitable to be treated as transactions. Assume that the business analyst chooses the nested transactions which are shown in Fig. 2. The figure was generated from Fig. 1 by hand-drawing dashed-line rectangles over Fig. 1 to represent the nested transactions. Thus, full-line rectangles in Fig. 2 represent the trade activities, which were already shown in Fig. 1, while the dashed-line rectangles represent the multi-step trade transactions that have names ending with the string "_tx". Fig. 2 thus shows the following nested transaction:

- *priceAndEscrow_tx* includes the PriceAndEscrow activity.
- *transportProduct_tx* includes subtransactions *getTrRequirements_tx* (determining transport requirements, obtaining rail insurance and obtaining a transporter) and *doTransport_tx* (actual product transport).
  - *getTrRequirements_tx* subtransaction includes the GetTrRequirements, GetRailInsurance, and GetRailTransport activities
  - *doTransport_tx* subtransaction includes the DoTransport activity.
- *receiveAndFinalize_tx* finalizes the transaction by receiving the product and completing the payment.

Each trade (sub)transaction is encapsulated in a separate smart contract. If a trade transaction fails, recovery procedures must be executed in reverse order, starting with subtransactions. Each recovery procedure notifies participants of the failure, enabling them to release their resources. Thus, three smart contracts are generated, one for each of the trade (sub)transactions *priceAndEscrow_tx*, *transportProduct_tx*, and *receiveAndFinalize_tx*, wherein the trade transaction *transportProduct_tx* includes subtransactions *getDocs_tx* and *doTransport_tx*, as shown in Fig. 2.

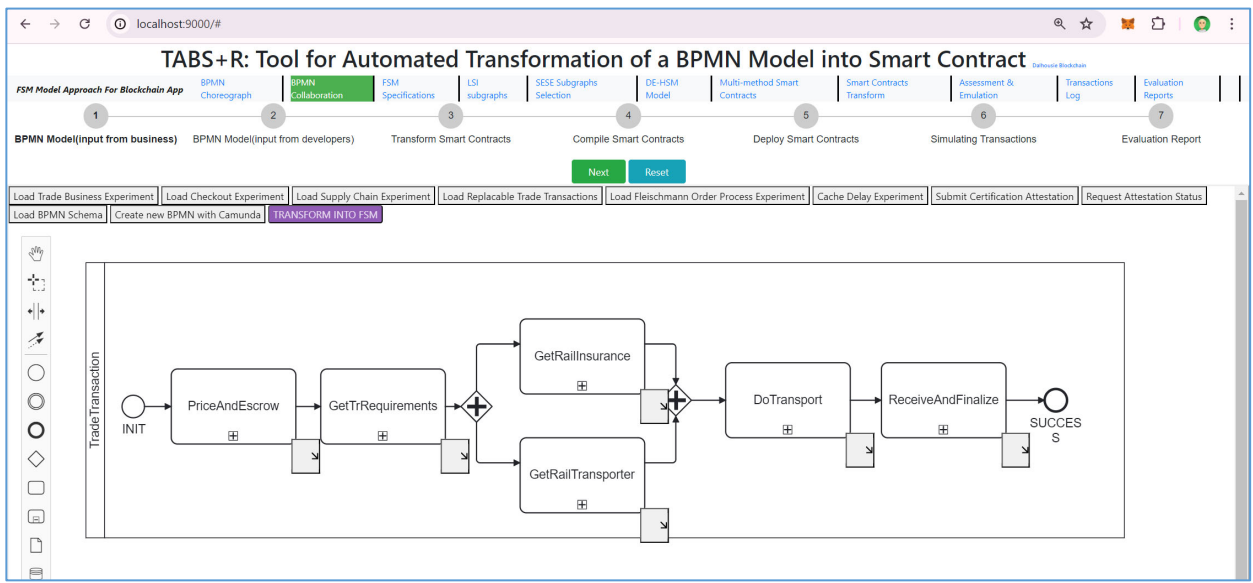

Fig. 1. Block diagram of trade activities represented using a business process model and notation (BPMN) model created using Camunda platform [11].

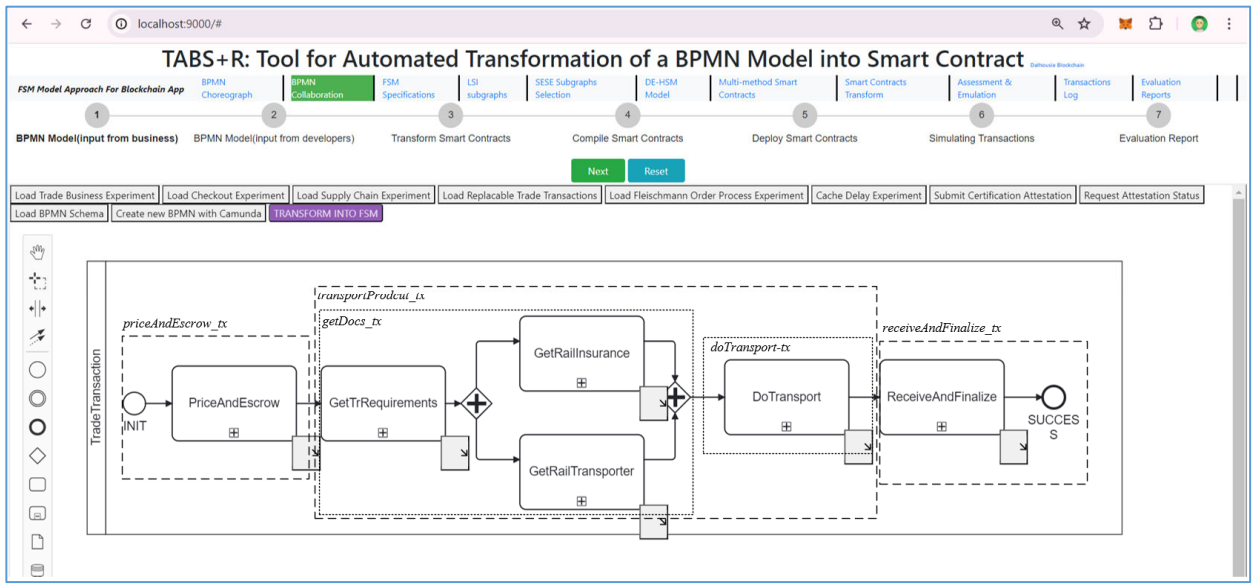

Fig. 2. BPMN model of trade activities as nested trade transactions.

In addition to the transactions shown in Fig. 2, there is an additional smart contract, referred to as the main smart contract, that includes all nontrade-transaction methods that invoke the trade transaction methods. Furthermore, the methods of a subtransaction are invoked from its parent transaction, while the methods of a trade transaction that is not a subtransaction are invoked from the main contract that contains all nontrade-transaction methods. In case of failure, recovery procedures for trade subtransactions are invoked in the reverse order of the first invocation of their methods. Each recovery procedure for a trade transaction produces events to notify each of the transaction participants about the failure.

4. Trade transaction upgrade and repair

Developers strive to anticipate potential issues that may arise during the execution of trade transactions and write exception handlers to manage them, trying to ensure successful completion of trade activities. However, not all failures can be anticipated. For instance, a flood washing out a railway line might prevent product transport, a situation unlikely to have been foreseen by the developer.

In such cases, when an unanticipated and uncaught exception occurs, the question arises about how to complete the trade activities when a part of the smart contract fails. Given the immutability of blockchains, representing new arrangements in the smart contracts is challenging. One approach could involve creating a new smart contract tied to the failed one while successfully leveraging completed activities. Alternatively, a new contract could be derived from the failed one, incorporating completed activities. In the following section, we describe our approach to facilitating repair to successfully complete the trade activity.

When a trade activity represented by a smart contract fails, the main objective is to make alternative arrangements that will overcome the failure, and then amend or repair the smart contract with the alternative arrangements to ensure the successful completion of the trade. Given that the smart contract is initially developed from a BPMN model of the trade activity, the repair process also involves creating a new BPMN model. This model represents alternative arrangements by modifying the original failed BPMN model, specifically replacing the pattern that caused the failure with one that includes alternative BPMN patterns that will not fail.

Our approach, to repairing trade (sub)transactions in smart contracts to reflect alternative arrangements, utilizes nested trade transactions to facilitate repair, focusing on amending the failed subtransaction rather than the entire trade activity. We automate the generation and deployment of smart contracts and describe how to update the failed subtransactions to ensure continued execution of the smart contract post-repair. Recovery from the failure of a trade (sub)transaction follows a well-defined process, as outlined in the previous section. Notably, we package and deploy each trade (sub)transaction as a separate smart contract, localizing the repair. The repair can thus be achieved by upgrading or replacing the failed (sub)transaction with a corrected version.

Failure is first analyzed to determine a BPMN pattern that corresponds to the innermost trade transaction in which the failure occurred. Repair is attempted within the context of the BPMN model that corresponds to the trade transaction first. If the repair succeeds, then it is localized to the innermost transaction. If the developer is unable to complete repair within the identified BPMN pattern that corresponds to the trade transaction, then the repair of that pattern is aborted and restarted, but within the context of the BPMN pattern of the parent trade transaction. Once the BPMN pattern is repaired, automated transformation of a BPMN pattern into a smart contract is used to generate the smart contract representing the repaired BPMN pattern.

We first outline the repair of the failed BPMN pattern. We then describe the generation of the smart contract for the repaired pattern, and how we achieve replacement, i.e., an upgrade of the failed smart contract with the repaired version.

## 4.1. Repair at the BPMN model level

The developer is presented with the original BPMN pattern that led to the failure, including the failure's cause. Then, the developer must replace this failure pattern with a new one that presumably avoids the failure. The repaired BPMN model integrates the successfully completed activities from the failed smart contract's execution along with new elements to complete the trade activity. This process, though abstracted, is outlined in Fig. 3.

The initial step in Fig. 3 involves determining which BPMN patterns within the model caused the failure. The cause of the failure is often due to an unhandled exception or an exception handler failing to resolve the issue. Since failures occur during the execution of a smart contract method, translating this failure information from the smart contract context to the context of the BPMN model is crucial. The developer uses this information to amend the BPMN pattern and repair the trade activity.

First, the failure information must be translated or mapped, from the context of a failed smart contract to the BPNM level representation, to provide the developer information about the BPMN pattern that needs to be amended or repaired and the reasons for failure. The developer is then presented with a BPMN pattern to be repaired and information about that pattern, including information flowing in and out of the pattern, purpose, and cause of the failure. The developer replaces the BPMN pattern causing the failure with a repaired BPMN pattern that is transformed into the methods of a smart contract that replaces its failed version.

---

1. **BPMN model failure information:** Information about the failure is gathered, identifying the BPMN pattern that caused it. The repair begins with the BPMN pattern associated with the innermost trade (sub)transaction where the failure occurred.
2. **Model amendment:** The developer is shown the original failing BPMN pattern ($Pf$), including details on the reason for failure, the pattern's intended function, and the objects and information involved. The goal is to replace $Pf$ with a repaired pattern ($Pr$) under the following constraints:
    (a) Pre-repair condition: The computation in the new pattern uses the same objects that were input to the failed execution.
    (b) Post-repair condition: The output objects from the new pattern must include, at a minimum, those produced by the failed computation.
    If the pattern cannot be amended while satisfying the above constraints, the repair escalates to the parent trade transaction's BPMN pattern.
3. **Smart contract generation:** Upon completing the BPMN model's repair, the system generates a new smart contract from the updated BPMN model of the pattern.

Fig. 3. Repair steps.

---

Consider a scenario where the *doTransport_tx* fails due to a washed-out rail line. The BPMN model shows to the developer information that insurance and a transporter had been arranged, but the transport could not occur. If an alternative route is available with the same transporter and insurance, the constraints are satisfied, and the repair remains within the *doTransport_tx* context. However, if the transport must switch to a road route with different insurance and/or transporter, the repair must "backtrack" to the parent transaction that also includes the GetRailInsurance and GetRailTransporter activities. If the repaired pattern does not meet the required outputs for subsequent activities, the repair extends to these activities, as outlined in Fig. 3. The final step involves generating a new version of the smart contract from the repaired BPMN model. Of course, before the repair of the smart contract can proceed, alternative transport arrangements need to have been discovered and arranged, as the BPMN model and the generated smart contract only carry out the actual trade activities.

Continuing with our example, assume that an alternative arrangement is found for the product transport, with the same transporter using an alternative rail-line route and existing insurance applied to the new route. Since the alternative route arrangements are performed by the same transporter and the insurance covers the transport via the alternative route, the already completed activities performed by the GetInsurance and GetTransporter sub activities do not need to be repaired. This is the case when the pre-repair condition (a) of step 2 in the repair outline shown in Fig. 3 is satisfied.

However, consider a situation where *doTransport_tx* fails, and no other rail route is available, necessitating road transport. Assuming that the hired rail transporter does not provide road transport, or the insurance for the rail transport does not apply to the road transport, then the pre-repair constraint (a) in step 2 is not met. Hence, the previously completed trade activities of obtaining insurance and arranging transport must also be repaired. In such a situation, the repair within the context of the innermost trade subtransaction is aborted, and the repair of the previously completed activities must be amended to accommodate alternative transport arrangements. In such a situation, the repair escalates to the parent transaction. Thus, the repair restarts for the BPMN pattern representing the parent trade transaction, *transportProduct_tx*. Since the *transportProduct_tx* parent transaction includes the GetInsurance and GetTransporter activities and the failed *transportProduct_tx* subtransactions, the developer must amend the BPMN pattern, including the GetInsurance, GetTransporter, and DoTransport trade activities.

When the computation flow exits the repaired BPMN pattern, it must produce information required by succeeding trade activities. If the repaired pattern produces all necessary information, the succeeding activities are unaffected and need no amendment. However, if the repaired pattern does not provide all information required for the subsequent activities, repair extends to the parent trade (sub)transaction. This represents the situation where the post-repair condition (b) of step 2 in Fig. 3 is not satisfied. The final step uses the amended BPMN model to generate smart contract methods for the repaired BPMN model.

In the following subsections, we elaborate on how the repair steps are accomplished within the context of automated smart contract generation from BPMN models. We also discuss issues that need to be addressed to bring automated generation of smart contracts from BPMN models closer to adoption for supporting trade activities. The PoC is described in the next section.

4.2. Repair: From the BPMN model to smart contract

In this subsection, we detail each repair process step shown in Fig. 3, focusing on transformations between the BPMN model and smart contract abstraction levels.

*4.2.1. BPMN model failure information*

Recall that the process of generating a smart contract from a BPMN model starts with analyzing the DAG representation of the BPMN model to find SESE subgraphs of the DE-HSM model, which the developer uses to define nested trade transactions. The DE-HSM model transformation into smart contract methods involves packaging and deploying each trade (sub)transaction as a separate smart contract with its methods.

Additionally, recall that failure is detected during the original smart contract's execution when an unhandled exception occurs. Since a trade (sub)transaction is defined using a SESE subgraph, identifying the smart contract where the exception was raised is straightforward when a failure occurs. The exception is raised in a method belonging to a trade transaction represented by a subgraph in a DE-HSM model that is built from a BPMN model's DAG representation. Consequently, the repair system can determine the BPMN pattern needing repair, corresponding to the SESE subgraph representing the trade transaction where the failure occurred.

*4.2.2. Model amendment*

After presenting the BPMN model and failure information to the developer, the developer attempts to repair the identified, failed BPMN pattern to generate a new, repaired BPMN pattern. The repair must satisfy the following constraints:

(a) Pre-repair condition: Information flowing into the repaired BPMN pattern *Pr* and its corresponding subgraph must match or be a subset of the information flowing into the original BPMN pattern *Pf*.

(b) Post-repair condition: Information flowing out of the repaired BPMN pattern *Pr* and its corresponding SESE subgraph must match or be a superset of the information flowing out of the failed BPMN pattern *Pf*.

These constraints ensure that the effects of executing the repaired BPMN pattern and its corresponding SESE subgraph and trade (sub)transaction are localized, not affecting preceding or succeeding trading activities. Thus, it must be ensured that information required for executing the pattern being repaired is produced by previously completed activities as per the pre-repair constraint. In our example, if the same transporter and insurance can be used for the alternative transport route, the repair can be accomplished within the *doTransport_tx* transaction context. The developer performing the repair determines whether the existing insurance and transporter are applicable for the repaired activities. If a new transporter and/or insurance are required, the repair process restarts for the parent transaction.

The post-repair constraint ensures that the repaired pattern's computation produces the information required for subsequent computations. This means that the objects and information flowing out of the repaired BPMN pattern *Pr* must include those flowing out of the failed pattern *Pf*.

*4.2.3. Repaired smart contract generation*

Each trade (sub)transaction is packaged and deployed in a separate smart contract. To facilitate replacing the failed smart contract with a repaired one generated from the repaired BPMN pattern with alternative transport arrangements, two tasks must be accomplished:

i. A new version of the smart contract generated from the repaired BPMN pattern needs creation and deployment.

ii. Invocation of the new version of the smart contract, representing the repaired pattern *Pr*, must replace the invocation of the original smart contract representing the failed BPMN pattern *Pf*.

For (i), the developer creates a new BPMN pattern for the failed activity, not the entire trade activity.

For (ii), the architecture of the execution model for smart contracts generated from BPMN models is exploited. Applications do not directly invoke smart contract methods; instead, they invoke an application programming interface (API) prepared by the TABS+R tool, which in turn invokes the smart contract methods. To "upgrade/repair" the trade (sub)transaction such as *transportProduct_tx*, the API is updated to point to the new smart contract version and its methods, ensuring the invocation of the new *transportProduct_tx* transaction version that replaces the failed version.

## 4.3. Discussion

Before repairing a BPMN pattern, alternative arrangements must be found by a business analyst or the developer, a process outside this paper's scope. However, once such arrangements are made, their representation in the BPMN pattern under repair is supported by our repair system, providing the developer with information on the objects flowing into and out of the computation performed by a trade (sub)transaction. Although we are progressing toward supporting smart contract generation for trade transactions and creating infrastructure for the automated generation and repair of trade transactions in the trade of goods and services, there are still aspects of utilizing the concept of nested transactions that need further investigation that we discuss below.

*4.3.1. Evaluating the pre- and post-repair conditions*

Although the developer has information on the objects and information flowing into the BPMN fragment to be repaired, currently the decision whether the repaired fragment satisfies the pre- and post-repair conditions is left to be made by the developer. Clearly, further assistance should be provided to the developer. For instance, insurance is required for the rail transport. If the rail line is washed out and the road transport is needed instead, how can it be recognized in automated fashion that the insurance for the rail does not apply to the road transport? Currently, we do not assist the developer in making such determination.

*4.3.2. Selection of nested transactions and their overhead*

Nested trade transactions incurring high overhead as an atomic commitment of subtransactions, within a parent transaction, is coordinated via the 2-Phase Commit (2PC) protocol, which requires the order of $n^2$ messages for $n$ participants. Consequently, the decision by the developer to deploy BPMN SESE subgraphs as nested subtransactions in the smart contract should not be made lightly. For instance, if the activities of a SESE subgraph manipulate only digital assets in such a way that a subtransaction activities can be easily compensated, then the use of subtransactions may be avoided by the developer carefully orchestrating compensating activities in the case of exceptions, which of course incurs the one-time overhead cost due to the developer's time to orchestrate the compensating activities. However, if the one-time task of writing code for the compensating activities by developers is high due to their complexity, automated generation of recovery procedures when nested subtransactions are used may be beneficial if the cost of the 2PC protocol for nested transactions is less than that of the developer time to write the compensating activities.

Another consideration is that the actual trade activities are recovered once they start. Consider, for instance, a simple case of ordering a number of parts that are to be assembled into a physical product under a deadline. First, parts are ordered, and when they arrive, the assembly and delivery of the product proceeds. The issue is that if the ordering of any part fails, then the product assembly fails due to missing the deadline. Consider, for example, the following two scenarios:

i. An order for a part may be canceled without any penalty.

ii. Once an order for a part is placed, canceling it incurs a penalty as the supplier initiates shipment via its sub-contractors immediately after an order is placed.

For (i) above, subtransactions are not required as placed orders for any parts can be canceled without any penalty through compensating activities represented by code produced by the developer. However, for the case (ii) the subtransactions are useful as using them facilitates placing the orders for all parts only after it is ascertained that all parts are available to order and hence avoiding the high cost of cancelling orders.

Thus, using nested transactions is cost effective if the cost of 2PC protocol execution within a smart contract is less than the cost of cancelling an order due to the suppliers' penalties plus the cost of the developer time to orchestrate compensating activities if canceling already placed orders for parts. Clearly further research on when to use nested transactions is needed.

5. Proof of concept: TABS+R tool

In our previous work [15], we demonstrated the feasibility of transforming BPMN models into smart contracts using the TABS tool. Subsequently, we introduced the concept of nested trade transactions in smart contracts to support complex collaborative activities beyond the capabilities of native blockchain transactions [22]. We extended the TABS tool into TABS+ to facilitate the automatic generation of smart contracts that implement these nested trade transactions while also providing a mechanism to support the transactional properties. To evaluate the feasibility of smart contract repair, we further augmented TABS+ into TABS+R, which supports the repair of smart contracts using the approach detailed in the preceding sections. We provide an overview of the tool's interface and outline the main steps involved in repairing a smart contract.

It should be noted that the TABS+R tool is configured for research, experimentation, and testing. It includes features and steps not intended for production environments, such as stepping through the transformation process and inspecting inputs and outputs. It also supports issuing multiple transactions and measuring various execution delays, by capturing timing at different points.

We first describe the process of transforming a BPMN model into smart contract methods using TABS+R, using the repair scenario from Figs. 1 and 2 as a case study. Next, we explain how the tool assists developers in facilitating repair and resuming trade activities.

5.1. Overview of the TABS+R tool

TABS+R utilizes the Camunda platform [28] for creating BPMN models according to BPMN specifications [26-27]. These models are stored in XML format. Fig. 1 displays a partial screen of TABS+R during BPMN model creation for our example application. Once the BPMN model is saved in XML format, it is transformed through a series of steps controlled via tabs in the tool's interface. The initial tabs focus on BPMN modeling, while subsequent tabs manage the transformation steps until the generation of smart contracts.

After creating the BPMN model, the developer guides its transformation into smart contracts by interacting with the tool and providing code for template methods that transformed for BPMN task elements. The tool supports the generation of smart contracts for Hyperledger Fabric blockchains or Ethereum virtual machine (EVM)-based blockchains, such as Quorum or Ethereum. The mainchain smart contract can invoke methods of smart contracts deployed on a sidechain.

Fig. 4 shows a screenshot of the BPMN model transformed into DE-HSM model with SESE subgraphs derived from the BPMN model of trade activities in Fig. 1. The left-hand panel displays the BPMN graph and SESE subgraphs identified by TABS+R. The right-hand panel lists these SESE subgraphs as selectable boxes, and the developer can instruct the system to deploy the selected subgraphs as trade (sub)transactions.

The developer decides which independent subgraph patterns should be deployed on a sidechain as separate smart contracts interacting with the main contract. The choice between Ethereum and Hyperledger Fabric for the mainchain and sidechain is made by the developer also. For testing, local blockchains are set up for both Ethereum and Hyperledger Fabric, with prepared channels on Hyperledger Fabric for smart contract deployment. For Ethereum-compatible sidechain, Quorum is used. After the selections, the model is transformed into methods of smart contracts that are then deployed and executed.

The developer can step through the system's execution, observing message-by-message progress. The tool graphically represents state changes in individual FSMs of the DE-FSM submodels (Fig. 5) to aid in testing and manual verification. Execution delays are displayed as the process proceeds. Additionally, a developer can generate exceptions for specific activities, which is useful for testing and evaluating transaction repair. In Fig. 5, fault exception propagation is highlighted in red.

Fig. 4. Identified single-entry single-exit (SESE) subgraphs and their selection.

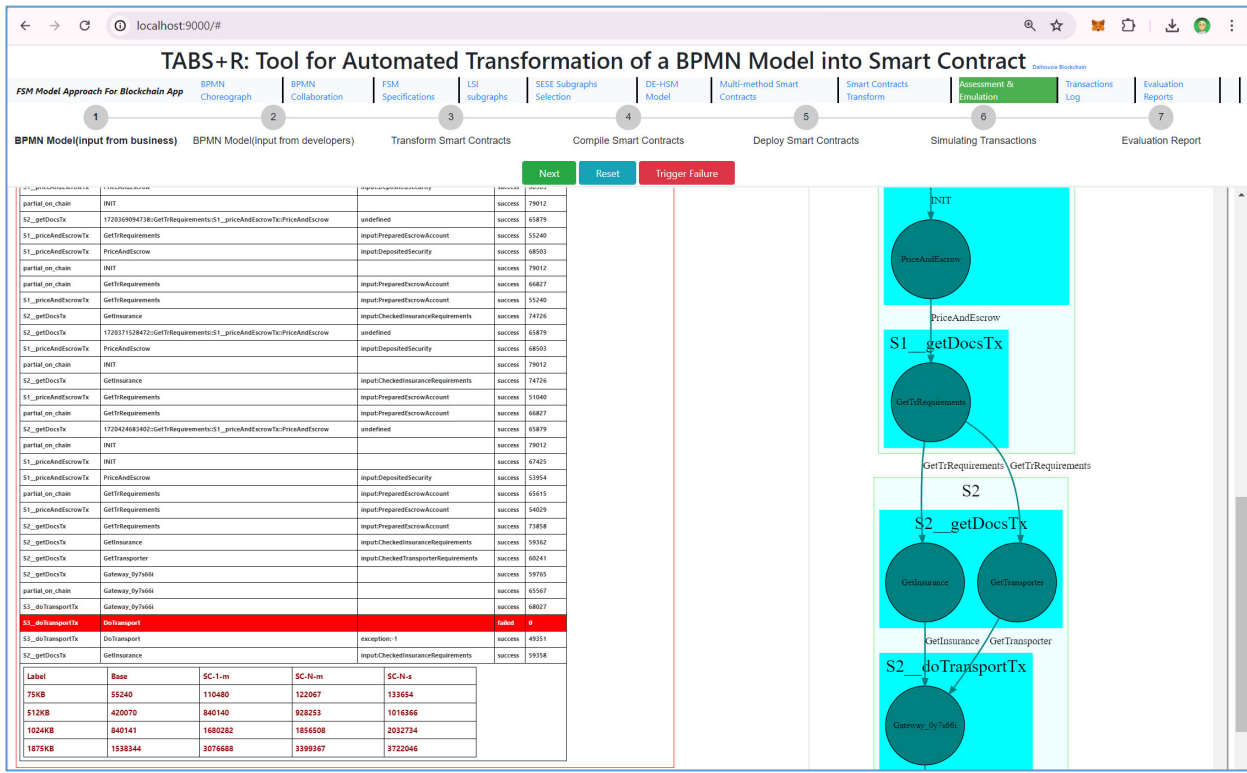

Fig. 5. Monitoring execution of trade (sub)transactions and their activities for evaluation purposes.

5.2. Repair of trade (sub)transactions with TABS+R

When an unhandled transaction failure occurs during the execution of a trade (sub)transaction, TABS+R presents the developer with a popup BPMN model fragment corresponding to the innermost subtransaction containing the failed trade activity. The developer then amends the BPMN model to create a repaired version that resolves the failure (Fig. 6).

In the repair process, the developer modifies the BPMN model of the failed trade (sub)transaction while ensuring that: (a) the input data flowing into the subtransaction, and (b) the output information flowing out, remain consistent with the previous failed version, i.e., the pre- and post-repair conditions are satisfied, and if not, the repair escalates to the parent transaction. Once the repair is achieved, a new smart contract for the repaired (sub)transaction is generated and deployed on the same blockchain as the original failed version. Additionally, the dApp's API must be updated to invoke the repaired subtransaction instead of the failed one. Once these changes are made, the developer instructs the tool to proceed with the execution of the repaired subtransaction.

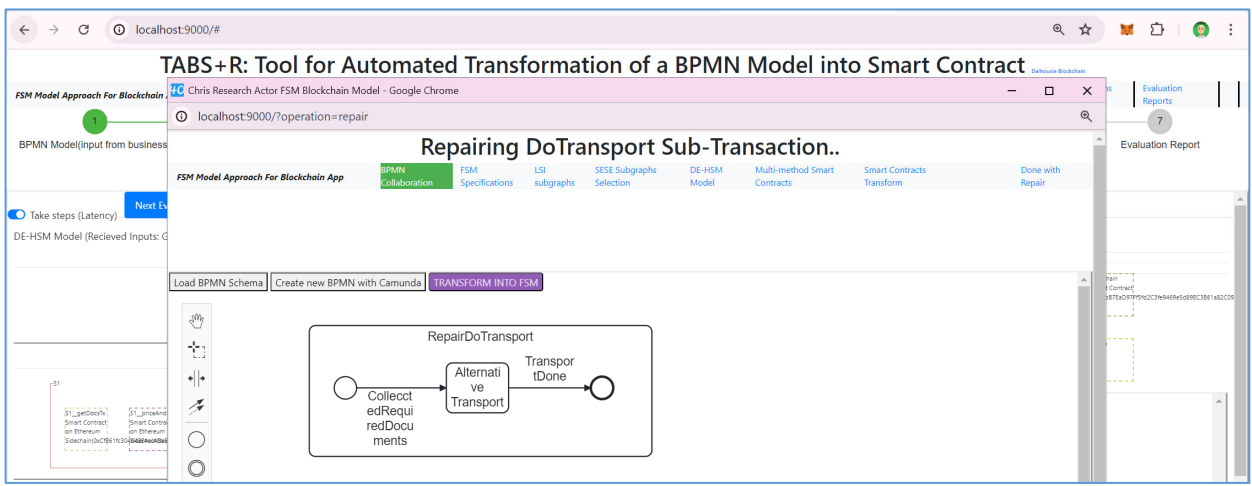

Fig. 6. Repair of the trade subtransaction *doTransport_tx*.

The achieved repair is contingent upon the applicability of input information, such as insurance documents and transporter selection, to the new transport method. For our example, if the same transporter is able to provide road transport instead of rail transport, and the original insurance also covers road transport, then repair of the *doTransport_tx* subtransaction proceeds as planned with the same insurance contract and the same transporter contract as for the original smart contract. However, if either of the two conditions is not met, the repair of the subtransaction is aborted. In such a case, the developer is led to repair the parent transaction instead, as shown in Fig. 7. This process involves arranging new insurance and transport by road, which need to be recorded in the smart contract. The whole BPMN model is shown after repair in Fig. 8, which shows the new transport requirements by road and new arrangements for insurance and transport by road.

After completing the BPMN model for the repair, the developer instructs the tool to transform the repaired model into a smart contract and deploy it. Finally, the developer directs the tool to execute the repaired subtransaction and continue the overall workflow execution.

5.3. Tool evaluation and developer feedback

Although the primary goal is to gain acceptance of the TABS+R tool among developers, the current version is not yet ready for formal evaluation. This is due to the tool's limitations and its interface, which is currently geared toward design and testing rather than production use. Additionally, attracting developers to test the tool without compensation has been challenging. However, informal demonstrations to a blockchain company developers have elicited favorable feedback and suggestions for improvement, indicating a positive reception of the tool and its approach. Future work will focus on overcoming existing limitations and enhancing the user interface according to user interface and user experience guidelines.

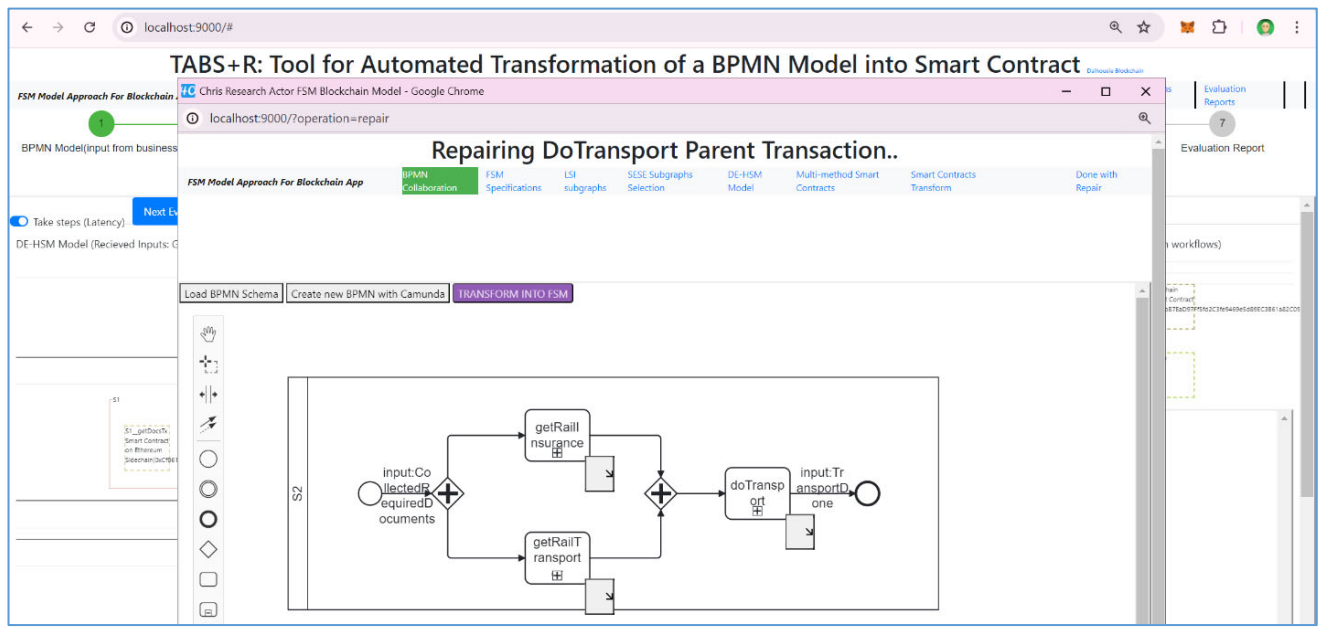

Fig. 7. Repair of the trade transaction *transportProduct_tx*, parent transaction of the failed *doTransport_tx*.

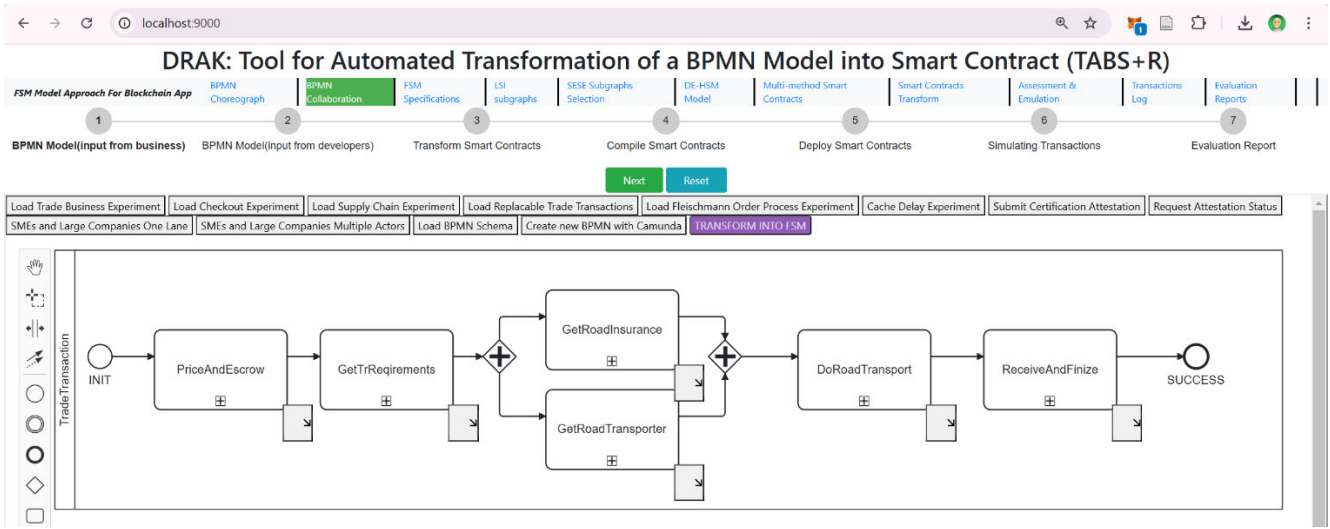

Fig. 8. BPMN diagram after the repair of the trade transaction *transportProduct_tx*, parent transaction of the failed *doTransport_tx*.

## 6. Related work, limitations, and future work

### 6.1. Related work

The Lorikeet project [12] uses a 2-phase approach to transform BPMN models into smart contracts. In the first phase, the BPMN model is analyzed and transformed into smart contract methods, which are then deployed and executed on a blockchain, specifically Ethereum. An off-chain component facilitates the communication with the dApp. The actors exchange messages according to the BPMN model, with these exchanges being managed by the off-chain component. The smart contract includes a monitor that stores and enforces the model choreography, ensuring that message exchanges adhere to the predefined sequence. In addition, the project provides support for asset control, including both fungible and nonfungible assets. It provides a registry of tokens and methods for asset management, such as transfers, thus enabling rapid prototyping of smart contracts from BPMN models that require such features. This allows for the creation, testing, and modification of smart contracts before deployment.

Caterpillar [11, 34] offers a different approach, focusing on BPMN models within a single pool, which is a BPMN construct, where all business processes are recorded on the blockchain. Its architecture comprises three layers: web portal, off-chain runtime, and on-chain runtime. The on-chain runtime layer includes smart contracts for workflow control, interaction management, configuration, and process management, with Ethereum as the preferred blockchain.

Loukil et al. (2021) [14] introduced CoBuP, a collaborative business process execution architecture on blockchain. Unlike other approaches, CoBuP does not compile BPMN models directly into smart contracts but instead deploys a generic smart contract that invokes predefined functions. It features a three-layer architecture: conceptual, data, and flow layers, transforming BPMN models into a JSON workflow model. This model governs the execution of process instances, which interact with data structures on the blockchain.

Di Ciccio et al. (2019) [35] compared the approaches of Lorikeet and Caterpillar across several features, including model execution, BPMN element coverage, incorrect behavior discovery, sequence enforcement, participant selection, access control, and asset control. They highlighted the unique aspects of each approach and provided a basis for comparison.

In Ref. [22], we expanded the comparison performed in DiCiccio by including of CoBuP [14] and TABS+ approaches and adding the following features for comparison purposes: support for nested transactions, deployment capabilities, type of synchronization, and privacy. The features that distinguish our approach and the TABS+ tool from others include the transformation of a BPMN into a DAG representation and then to the DE-HSM and DE-FSM models, which enable the analysis of process flow to identify localized processing using SESE subgraphs. Unlike other systems that use direct transformation of BPMN models into smart contract code, TABS+ produces smart contracts that are abstract representations of process flows with details expressed through concurrent FSMs. The logic of the process is executed by a smart contract executing the firings of concurrent FSMs, while some of the transitions cause executions of localized BPMN tasks, each task executed by a smart contract method with well-defined inputs and outputs. Furthermore, localized SESE graphs may be packaged and deployed as separate smart contracts that may be deployed on sidechains for efficiency, or they may be used to define nested trade transactions with well-defined transactional properties.

Similar to CoBuP, Bagozi et al. [36] used a three-layer but simpler approach. In the first layer, the collaborative process is represented in the BPMN by a business analyst. In the second layer, a business expert adds annotations to the BPMN model that identify trust-demanding objects, and then abstract smart contracts, which are independent from any blockchain technology, are created. Finally, smart contracts are created and deployed on a specific blockchain.

Mavridou and Lazska [37, 38] advocated the correct design of smart contracts. They target systems whose requirements are represented using an FSM that is transformed into the methods of a smart contract. Transformation may result in execution of specific tasks that are represented as methods of smart contracts. Then, each method of the smart contract is analyzed and patched for any security holes. They develop a tool as a PoC, called FSolidM, that examines each smart contract method and augments it with security codes to eliminate discovered security issues. Furthermore, after hardening a smart contract method, steps are taken to prevent the developer from modifying the inserted security code when amending the method's functionality before the smart contract deployment.

Mavridou et al. [39] advanced the work on FSolidM by creating a framework, called VeriSolid [39]. The smart contract is specified via a graphically specified transition system, which is an FSM extended with variable declarations and guards expressed in the Solidity language, plus a list of verifiable properties or constraints. The transition system is analyzed using the declarations and guards, and is transformed into a transition system with augmented states to ensure that the desirable properties are satisfied. Only then is the system transformed into the methods of a smart contract expressed in Solidity, resulting in a correct design in terms of satisfying the verifiable properties.

In our approach, once the BPMN model is transformed into the DE-FSM model, it represents the business logic using concurrent FSMs, which is a transition-based system. Similar to FSolidM, we check the methods written by the developer to represent BPMN tasks for known security bugs and prevent them. However, we do not yet support representation of the desirable properties in the BPMN model.

The verification of smart contracts generated from BPMN has been addressed in Ref. [40]. They use a two-step transformation process. In the first phase, the BPMN model is transformed to Prolog, which is used to validate the business logic. Following this, they transform the BPMN model into a smart contract in GO language. However, there is no validation of the generated code.

When smart contracts are used to represent the collaborative activities of participants, it has been acknowledged that the blockchains' immutability property is desirable for promoting trust, but it also causes difficulties, as frequently, such collaborations need to be augmented to either repair security issues or support new desirable features. Thus, the upgradability of smart contracts is an important issue that has been tackled in literature, particularly in the context of security of smart contracts, as is demonstrated by literature surveys that have been performed on this topic [50-52]. For instance, Rodler et al. [41] described a framework, EVMPatch, which uses a bytecode rewriting technique to automatically rewrite common off-the-shelf contracts to upgradable contracts. EVMPatch upgrades faulty Ethereum smart contracts using a bytecode rewriting technique to patch common bugs, such as integer over or underflows and access control errors.

Another bug fixing framework is SolSaviour [42] that uses a voteDestruct mechanism to allow contract stake holders to decide to withdraw inside assets and destroy the defective contract, which is followed by repairing and redeploying the smart contract while incorporating the old contracts' internal states into the new ones. Our tool TABS+R is similar when the repaired version of the smart contract incorporates the successfully completed activities of its unrepaired version.

Aroc [43] is yet another framework for repairing smart contracts written in Solidity for an EVM with the objective of not modifying the vulnerable smart contract itself. Instead of using a proxy contract that invokes a patched or repaired and redeployed version of the vulnerable smart contract, the authors propose a new smart contract that is created for preventing attacks on the faulty contract. The Aroc system first generates and deploys a patch smart contract that blocks invocations of the vulnerable smart contract by malicious smart contracts that try to exploit the vulnerability. The owner of the vulnerable smart contract uses a special transaction supported by an extended EVM that supports the redirection of the invocations of the vulnerable smart contract to the patch smart contract.

Corradini et al. [44, 45] address the tension between the trust in blockchain, achieved via immutability, and the need for flexibility, which is required for multiparty collaboration. They use BPMN to model business processes, which are then transformed into code, while it is the code's execution state that is stored on the blockchain. This decoupling of business process choreography from its execution state allows for run-time changes to the process execution. They developed a tool, FlexChain, to show feasibility of their approach.

Falazi et al. [46] addressed the issue of using smart contracts in support of business process modeling and developed a prototype, BlockME, to validate their approach. They use BPMN to represent the choreography of the business processes while extending BPMN task to support invocation of smart contract methods for permissionless blockchains. The BPMN model is transformed into BPEL for execution. The choreography of processes is executed via BPEL that invokes smart contract methods that implement the BPMN task elements. Authors extend their approach and BlockME prototype in Ref. [47] to support invocation of smart contract functions of different blockchains, including both permissioned and permissionless by developing a new technique to identify smart contract functions and a metric to gauge transaction finality. In short, the collaboration logic is executed in BPEL and not like the smart contract methods in our TABS+ approach.

The closest work to our approach in upgrading smart contracts is Ref. [48]. This paper analyzes and implements three different upgradeability concepts, one based on a registry, one using a proxy pattern, and the third one based on a registry combined with a pattern segregation. The case they use is a large organization with departments. The BPMN model of the collaboration is transformed into smart contracts, with each department's activities represented by a smart contract. The upgrade thus needs to be carefully managed to ensure that activities already executed are consistent in the context of the upgrade. Their findings suggest that the unstructured storage proxy pattern is the most promising for practical use, especially regarding cost-effectiveness and minimal added complexity.

In comparison to Ref. [48], our TABS+R approach naturally packages activities into different smart contracts, wherein our use of nested trade transactions is exploited to protect against inconsistencies due to some of the activities being completed by the old version of the smart contract while some are executed with the newly upgraded smart contract.

To the best of our knowledge, we are not aware of any formal work on multistep and multimethod transactions for blockchain smart contracts, to which we refer simply as trade transactions, wherein a trade transaction contains executions of independently invoked smart contract methods by different actors. We introduced the concept of a multimethod transaction in [15, 22], and in this paper, we exploited it to upgrade smart contracts in the context of automated generation of smart contracts from BPMN models.

### 6.2. Limitations and future work

Although we feel that our approach to ease the developer's task in creating smart contracts for trade transactions is progressing, there are still many problems and limitations that need to be addressed. In this subsection, we describe both the limitations and our plans on how to address them. These are besides the issues of the high cost of nested transactions and the determination of the pre- and post-repair conditions that were discussed in Section 4.3.

#### 6.2.1. Securing smart contract methods

To secure smart contracts, we adopted the approach in [37, 38], wherein the authors propose the hardening of smart contracts created by transformation of an FSM to smart contract methods. Given that an FSM is a representation of the smart contract activities, they proposed a transformation of the FSM into the methods of a smart contract. They then propose securing each of the smart contract methods by inserting security patterns to guard it against (i) re-entrancy issues, (ii) transaction ordering in the face of unpredictable states, (iii) timed transitions, and (iv) access control. We successfully incorporated the re-entrancy protection into TABS+ by inserting appropriate locking patterns into the smart contract and supporting access control. In essence, the security patterns are inserted into the smart contract method at the start of a method and its end. However, in our future work we shall develop smart contract patterns to guard against all known smart contract vulnerabilities. In addition, unlike native blockchain transactions, for trade transactions we also need to protect against man-in-the-middle attacks.

#### 6.2.2. Validation and verification

Validation and verification need to be an integral part of the transformation process from BPMN models to smart contracts deployment. Transformations of BPMN models result in a DE-FSM model that is a transition system, and we plan to apply the VeriSolid [39] verification methods to ensure that generated smart contracts are correct by design. We already ask the BPMN modeler to document information flowing along with the execution flow. We shall also extend that documentation with the desirable properties and then perform formal verification of the system in terms of liveliness, reachability, and deadlock-free properties, so that the desirable properties are satisfied.

#### 6.2.3. Blockchain agnostic smart contracts

One of our objectives is to achieve the generation of smart contracts that are blockchain agnostic. We made progress toward this objective by representing the collaboration in a blockchain independent manner by expressing it in terms of the interconnection of the DE-FSM models. However, currently, to apply a smart contract developed for one blockchain to be deployable and executable on another blockchain, the scripts for methods representing the BPMN task elements are provided by the developer and need to be executable on the target blockchain. To overcome this issue, we are investigating a two-layer approach taken by the Plasma project, described in Ref. [49], in which the task scripts are not executed on the blockchain but rather off-chain, while the smart contract simply guides the collaborations and obtains certifications about the results of the tasks that are executed off-chain.

## 7. Summary and conclusions

The trade of goods and services, including distributed finance, contains activities that require specialized customization that is not yet easy to support by traditional development of smart contracts. Activities in the trade of goods and services are subject to effects from external events that may cause failure of a trade activity supported by a smart contract. Such a failure thus precludes successful completion of a smart contract unless that failed activity can be replaced with one that will succeed and facilitate successful completion of the trade activity. We call such a upgrade of a smart contract as a transaction repair, although others may use the terms upgrade or replacement. We described how we use the concept of nested trade transactions, together with automatically generated transaction mechanism, to support the repair or upgrade of a failed smart contract so that it can be completed. The repair process exploits the concepts of the nested trade transactions to ensure that the successfully completed activities of the failed smart contract version may be incorporated consistently into the new version of the smart contract, modeling the alternative trade activities to avoid failure.

However, when a trade activity needs to be repaired to respond to external situations, such repairs are performed by business analysts who can then repair the BPMN model representing the activity. As the trade activity is represented by a smart contract, replacing an activity in a smart contract requires an effort that is equivalent to writing a smart contract in the first place. This is because the developer's time needs to be allocated to the activity and the developer must familiarize herself/himself with the requirements of the contract and the changes that are required, write and test the amendment to the smart contract, and deploy the smart contract.

As can be appreciated, delays with allocating a developer's time, and developer's time needed to write the new version of the smart contract replacing the failed transactions are too long, particularly in situations when the repair of a smart contract representing trade activities needs to be made promptly, such as the use case described in this paper. Thus, we strive for fully automated generation of smart contracts and their repair that can be under the control of a business analyst only, without assistance of a software developer.

There are two obstacles in that effort. Software developer is required for the selection of which SESE subgraphs should be deployed as trade transactions. This leads to difficulties already discussed in the subsection IV. The major obstacle, however, is the generating methods that implement the PBMN task elements, which currently need to be provided by the developer. We are investigating if there is some simple language or model to represent the BPMN task functionality that can be automatically translated into a smart contract method for the target blockchain.

By supporting automated creation of smart contracts from BPMN models and providing support for augmentation of BPMN models with BPMN patterns and replacement of patterns in BPMN models with similar patterns, we are striving to create an environment to provide a relatively new concept of smart-contract-as-a-service (SC-as-a-service). In short, a modeler would be able to search a repository for BPMN models for major activities, such as a letter of credit, and customize the BPMN model by replacing patterns representing transactions or subtransactions, with similar patterns for customization purposes to suit the specific context, and then use the TABS+R tool to transform the BPMN model into a smart contract and deploy it on the blockchain in automated fashion.

## 2. Conflict of interest

The authors declare no conflict of interest.

## 3. Author contributions

All three authors participated in research and writing of this paper.